\documentclass[letter,twocolumn]{jpsj3}
\usepackage{txfonts}
\usepackage{graphicx}% Include figure files
\newcommand{\vc}[1]{{\boldsymbol{#1}}}
\newcommand{\e}{\varepsilon}

\title{Electrical and Thermoelectrical Transport Properties of
Dirac Fermions through a Quantum Dot}

\author{Tomosuke Aono}
\inst{Department of Electrical and Electronic Engineering,
Ibaraki University,
Hitachi 316-8511, Japan} %\\

\abst{We investigate the conductance and thermopower
of massless Dirac fermions through a quantum dot
using a pseudogap Anderson model in a noncrossing approximation.
When the Fermi level is at the Dirac point,
the conductance has a cusp where the thermopower changes its sign.
When the Fermi level is away from the Dirac point,
the Kondo temperature shows a quantum impurity transition between
an asymmetric strong-coupling Kondo state and a localized moment state.
The conductance shows a peak near this transition and reaches the unitary limit
at low temperatures.
The magnitude of the thermopower exceeds $k_B/e$, and
the thermoelectric figure of merit exceeds unity.}

\kword{Dirac fermions, pseudogap Kondo effect, electron transport, thermoelectrics, quantum dot}

\begin{document}
\maketitle

Electron transport
in graphene~\cite{Neto:2009cl} 
is currently under active investigation.
Graphene-based quantum dot (QD) structures are fabricated and
Coulomb blockade peaks are observed~\cite{%
Stampfer:2008dq,Ponomarenko:2008ej,
2009NanoL...9.2891M,
Guttinger:2010in}.
In semiconductor QD systems, it is known that 
the exchange coupling between a local spin in a QD
and conduction electrons 
induces the Kondo effect~\cite{1998Natur.391..156G,1998Sci...281..540C}.
The Kondo effect in graphene has been discussed
in connection with the magnetic impurity problem in massless Dirac fermions~\cite{%
Sengupta:2008gf,
Uchoa:2009hz,Cornaglia:2009ir,Zhuang:2009ji,
Wehling:2010in,Vojta:2010dw,
Uchoa:2011gt,
%Uchoa:2011fr,
2012arXiv1208.3113F}.
A recent experiment shows that the Kondo effect can be induced by lattice vacancies~\cite{Chen:2011jm}.
The Kondo effect probably occurs
in another representative Dirac fermions, that is, the surface state of a topological insulator.
A single Dirac cone appears
on the surfaces of three-dimensional topological insulators~\cite{Xia:2009fn,Hsieh:2009dp,Hsieh:2009ig,Chen:2009do,Zhang:2009ks}.
Several theoretical studies on the Kondo effect in this system have been conducted~\cite{Liu:2009ig,Tran:2010dw,Zitko:2010bp,Feng:2010jx,2012arXiv1211.0034M}.

%--------------------------------------------
The Kondo problem in massless Dirac fermions
is an important part of
the pseudogap Kondo problem~\cite{Withoff:1990ce},
in which
the density of states of
conduction electrons $\rho(\omega)$ obeys
a power law:
$\rho(\omega) \propto |\omega|^r$.
For example, it is realized in
unconventional superconductors~\cite{Withoff:1990ce,Cassanello:1996jg
%Cassanello:1997ku
}.
The numerical renormalization group (NRG) calculations~\cite{GonzalezBuxton:1996cz,
%GonzalezBuxton:1998hb,
Bulla:1997fp} and
perturbative scaling theory~\cite{Fritz:2004cn,Vojta:2004gn}
show that the system exhibits
an impurity quantum phase transition.
For the massless Dirac fermion model, in which $r=1$,
when the Coulomb interaction is sufficiently strong,
there are three fixed points:
the local moment (LM) fixed point,
the asymmetric strong-coupling (ASC) or frozen impurity fixed point, 
and the valence fluctuation (VFI) fixed point located in between~\cite{
%GonzalezBuxton:1998hb,
GonzalezBuxton:1996cz,
Fritz:2004cn,Vojta:2004gn}.
The Kondo problem in graphene has been studied 
as a tunable pseudogap Kondo problem
\cite{Sengupta:2008gf,Vojta:2010dw},
where the transition can be controlled by external gate voltages.
Recently, the Kondo effect indicated in ref.~\citen{Chen:2011jm}
has been analyzed using a pseudogap Anderson model~\cite{Kanao:2012ho}.

%--------------------------------------------
In addition to their electrical transport properties,
nanostructured materials have been
investigated to improve their thermoelectrical properties~\cite{Dubi:2011ko},
which also reveal the electronic states in the materials.
For instance, 
the measurement of the thermopower 
shows the electron-hole asymmetry in a system. 
The thermopower under the Kondo effect~\cite{Kondo:1965gu}
has been examined in QD systems theoretically~\cite{2001EL.....56..576B,Dong:2002by},
and
its measurement clarifies the formation of the Kondo resonant state~\cite{Scheibner:2005kh}.
Moreover,
typical topological insulators~\cite{Xia:2009fn,Hsieh:2009dp,Hsieh:2009ig,Chen:2009do,Zhang:2009ks}
have been known as good thermoelectrical materials.

%--------------------------------------------
We investigate electrical and thermoelectrical transport properties
of Dirac electrons
through a QD
via tunneling barriers.
The barriers exhibit pseudogaps, which 
result in an impurity quantum phase transition.
We discuss this transition
in connection with electron transport properties.
We also show that thermoelectrical properties are enhanced
by a pseudogap.
To this end, we study 
the conductance,
the Kondo temperature
defined by the impurity magnetic susceptibility,
the thermopower,
and the figure of merit,
using a pseudogap Anderson model
in a noncrossing approximation (NCA)~\cite{Coleman:1984fz,Bickers:1987dc,Wingreen:1994hw,Vojta:2001fo}.

\begin{figure}
\begin{center}
\includegraphics[width=7cm]{%
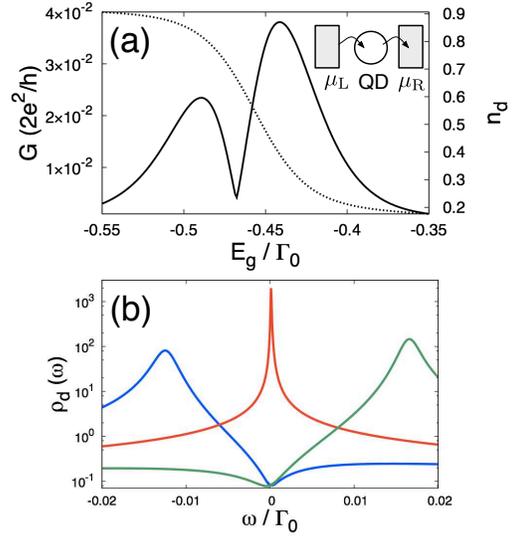}
\end{center}
\caption{%
(Color online)
(a) Conductance $G$ (solid line) and occupation number in the
 QD $n_d$ (dotted line) as functions of gate voltage $E_g$ for
$T/\Gamma_0= 1.0 \times 10^{-2}$
and $D/\Gamma_0=10$ $(\alpha=0.1)$.
Inset: schematic picture of the QD system.
(b) Local densities of states [$\rho_d(\omega)$] in units of $1/\Gamma_0$ at
 the left peak (blue), 
 cusp (red), and right peak (green) of $G$ in (a).
}
\label{fig:undoped}
\end{figure}

%--------------------------------------------
{\it Model---}
We consider a system
consisting of a QD connected to
the lead $i \;\;(i={\rm L,R})$
with a single Dirac cone and
the chemical potential $\mu_{i}$.
We discuss electron transport through the QD,
as shown in the inset of Fig.~\ref{fig:undoped}(a).
We focus on the zero-bias-voltage limit,
$\mu_{\rm L} \rightarrow \mu_{\rm R}$, setting $\mu_i=0$.
The position of the Dirac point from the Fermi level is denoted by $-\mu_0$.

%--------------------------------------------
The Hamiltonian of the lead $i$ 
is given by a Dirac Hamiltonian:
\begin{eqnarray} 
&& H_0^{(i)}  =
\int \frac{d^2k}{(2\pi)^2}
(\Psi^{\dagger}_{ai}(\vc{k}), \Psi^{\dagger}_{bi}(\vc{k})) M(\vc{k})
\begin{pmatrix}
\Psi_{ai}(\vc{k}) \\
\Psi_{bi}(\vc{k})
\end{pmatrix},
\label{eq: }
\end{eqnarray}
with
$
M(\vc{k})  =
\bigl(
\begin{smallmatrix}
-\mu_0 & \hbar v_{\textrm{F}} k e^{-i \theta} \\
\hbar v_{\textrm{F}} k e^{i \theta} & -\mu_0
\end{smallmatrix}
\bigr),
$
where
$\Psi_{ai}(\vc{k})$ are the annihilation operators of Dirac fermions,
$v_{\textrm{F}}$ is the Fermi velocity,
$\vc{k} = (k \cos \theta, k \sin \theta)$ with
$k = |\vc{k}|$, and $\theta$ stands for the azimuthal angle of $\vc{k}$.
The indexes $a$ and $b$ refer to (pseudo) spin indexes.
For a single-Dirac-cone system,
those are the spin indexes
$a=\uparrow$ and $b=\downarrow$.
For graphene,
$a=(A,s)$ and $b=(B,s)$ with the two sublattices $A$ and $B$,
which play the role of a pseudo spin,
and the spin index $s=\uparrow$, $\downarrow$ 
around the $K$ and $K'$ points.~\cite{note1}
%\footnote{For the $K'$ point, the off-diagonal elements of $M$ are
%replaced by their complex conjugates.}.

%--------------------------------------------
The QD has an energy level $E_{\rm g}$ controlled
by a gate voltage. 
The  Hamiltonian of the QD is given by
\begin{eqnarray}
H_{\rm d}  &=&
 \sum_{s=\uparrow,\downarrow} E_{\rm g} d^{\dagger}_{s} d_{s}
 + U n_{\uparrow} n_{\downarrow},
\label{eq:h_d}
\end{eqnarray}
where 
$d_{s}$ is the annihilation operator of an electron with spin $s$ in the QD,
$n_{s} = d^{\dagger}_{s} d_{s}$,
and $U$ is the Coulomb energy in the QD.
The tunneling Hamiltonian is given by
$
H_{\rm T}  = \sum_{i={\rm L,R}, s} \int \frac{d^2k}{(2\pi)^2} 
( 
V_i(\vc{k}) d^{\dagger}_{s} \Psi_{si}(\vc{k}) 
  + \textrm{h.c.})
$.
For simplicity, we assume $V_i(\vc{k}) = V$.

%--------------------------------------------
It has been shown that
the above model
reduces to the pseudogap Anderson model
after a series of linear transformations~\cite{Cassanello:1996jg,Sengupta:2008gf,Zitko:2010bp}.
In the energy space, $H_0$ is given by
\begin{eqnarray} 
H_0 &=& 
\sum_{i={\rm L,R} \atop \sigma=\pm}
\int d\omega \; \omega\; 
c^{\dagger}_{\sigma i}(\omega) c_{\sigma i}(\omega),
\label{eq:h_0}
\end{eqnarray}
where $c_{\sigma i }(\omega)$ is the annihilation operator of 
an electron in the lead $i$ with the pseudo-spin index $\sigma$ at the energy $\omega$.
The tunneling Hamiltonian $H_{\rm T}$ is given by 
\begin{eqnarray}
 H_{\rm T}  &=&  
  \sum_{i,\sigma} \int d\omega \;
  \sqrt{\Gamma(\omega)}
   d^{\dagger}_{\sigma} c_{\sigma i}(\omega) 
  + \textrm{h.c.},
  \label{eq:h_T}
\end{eqnarray}
where
$
\Gamma(\omega)   = \alpha |\omega + \mu_0|
$
with
$
\alpha  = \frac{V^2}{2\pi (\hbar v_{\rm F})^2}
$~\cite{Cornaglia:2009ir,Feng:2010jx,Zitko:2010bp}.
The power law behavior of $\Gamma(\omega)$ is terminated at
the band cutoff $D$~\cite{Withoff:1990ce,GonzalezBuxton:1996cz};
thus, it is convenient to
rewrite $\Gamma(\omega)$ in the form~\cite{GonzalezBuxton:1996cz}
\begin{eqnarray} 
\Gamma(\omega)  &=& \Gamma_0 
\left|  \frac{\omega+\mu_0}{D} \right|,
\label{eq:Gamma}
\end{eqnarray}
with $\Gamma_0/D =\alpha$.

%--------------------------------------------
We treat the infinite $U$ limit of
the model,
$H=H_{\rm d} + H_{\rm T} + H_0$,
which corresponds to
the asymmetric pseudogap Anderson model~\cite{%GonzalezBuxton:1998hb,
GonzalezBuxton:1996cz,Fritz:2004cn,Vojta:2004gn}.
We introduce the auxiliary operators,
$d_{\sigma}= b^{\dagger} f_{\sigma}$, 
where
$b$ is the boson operator and
$f_{\sigma}$ is the fermion operator.
The constraint
$b^{\dagger} b + \sum_{\sigma} f^{\dagger}_{\sigma} f_{\sigma}=1$ 
should be satisfied~\cite{%  
Barnes:1976tq,
%Barnes:1977um,
Coleman:1984fz}.
Then, we adopt the NCA~\cite{Coleman:1984fz,Bickers:1987dc,Wingreen:1994hw,Vojta:2001fo}
to calculate the local density of states on the QD and
other quantities.
To perform numerical calculations,
we evaluate Green's functions in the range of $|\omega| \leq 10 D$ and
introduce the Lorentzian cutoff in $\Gamma(\omega)$,
$\Gamma(\omega) \rightarrow \Gamma(\omega) \cdot D^2/(\omega^2+D^2)$.
The convergence of the NCA equations is monitored using
the sum rules on the boson and fermion Green's functions, 
and the sum rule on the total occupation number in the QD~\cite{Wingreen:1994hw}.
Those relations are satisfied within $0.1\%$.

The conductance $G$ and
the thermopower $S$
are given by 
$G=e^2 I_0(T)$ and $S=-I_1(T)/[e T I_0(T)]$ with
\begin{eqnarray}
I_n(T)  &=& -2/h \int d\omega \omega^n  \frac{\partial f(\omega)}{\partial \omega}
 \Gamma(\omega) {\rm Im} A^{r}(\omega),
\label{eq:integral}
\end{eqnarray} 
where $f(\e)$ is a Fermi-Dirac function,
$
 f(\omega) = 1/[\exp\left(\frac{\omega}{k_B T}\right) + 1],
$
and $A^{r}(\omega)$ is the retarded Green's function on the QD~\cite{Dong:2002by,2010PhRvB..81x5323L}.
The impurity magnetic susceptibility $\chi(\omega)$ is given by
$\chi(\omega) = \int_{-\infty}^{\infty} dt e^{i (\omega + i 0^{+}) t}
M(t)$, where
$ M(t) = i \theta(t) \langle [ \hat{M}(t),\hat{M}(0)] \rangle$
with
$\hat{M}= g \mu_B/2 \; 
(f^{\dagger}_{\uparrow} f_{\uparrow} - f^{\dagger}_{\downarrow}f_{\downarrow})$.
The static susceptibility $\chi= \chi(0)$ is evaluated using the NCA~\cite{Bickers:1987dc}.
We set $g\mu_B=1$ and $k_{\rm B}=1$ below.

%-------------------------------------------
% Figure: 
%         G and N_d
%-------------------------------------------

%--------------------------------------------
{\it Undoped system---}
Let us first consider $G$ and the dot occupation number
$n_d$ when the Fermi level is at the Dirac point: $\mu_0=0$.
It has been shown that, in this case, the Kondo temperature $T_K=0$~\cite{Withoff:1990ce,Sengupta:2008gf,Vojta:2010dw}.
In Fig.~\ref{fig:undoped}(a),
$G$ and $n_d$ are plotted 
as functions of the gate voltage $E_g$.
There is a cusp in $G$ at $E_{\rm g}=E_{\rm g}^*$.
Near the cusp, $G$ is linear in $|E_{\rm g}- E_{\rm g}^*|$, 
where
$n_d$ changes gradually.
This means that the system is in the VFI regime.
When $|E_{g}-E_{g}^*|$ increases further, $G$ decreases monotonically
and shows a double-peak structure.
%--------------------------------------------
In Fig.~\ref{fig:undoped}(b), 
the local density of states on the QD, $\rho_d(\omega)=(-1/\pi){\rm Im} A^r(\omega)$, is plotted. 
At $E_g=E_g^*$, $\rho_d(\omega)$ shows
a singular peak at the Fermi level, $\omega=0$,
which has already been discussed~\cite{Bulla:1997fp,Fritz:2004cn,Vojta:2010dw}.
In spite of this sharp peak,
the pseudogap in $\Gamma(\omega)$ minimizes $G$.

%-------------------------------------------
% Figure:
%    Temperature dep of G and S
%-------------------------------------------
\begin{figure}
\includegraphics[width=7cm]{%
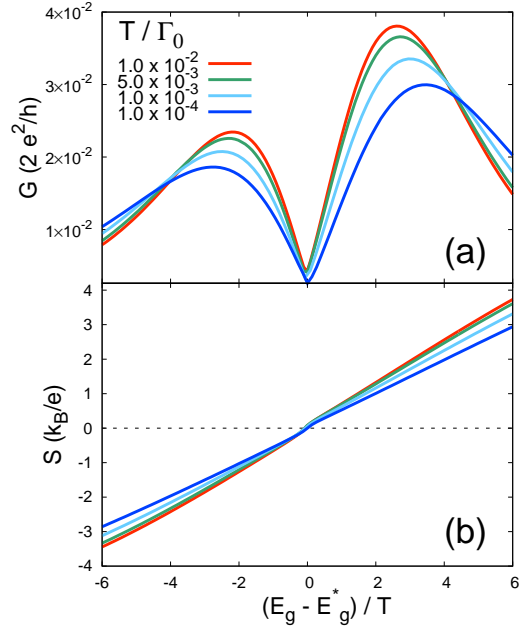}
\caption{%
(Color online)
Conductance $G$ (a) and thermopower $S$ (b) for $\mu_0=0$
as functions of normalized gate voltage, $(E_{\rm g}-E^{*}_{\rm g})/T$,
where $E_g^*$ is the gate voltage at the cusp of $G$.}
\label{fig:egdep}
\end{figure}

%--------------------------------------------
In Fig.~\ref{fig:egdep},
$G$ and the thermopower $S$ are plotted as functions of
$(E_g-E_g^*)/T$ at several temperatures.
In each figure,
the results are similar in shape.
This shows that the separation of the double peaks of $G$ depends linearly on $T$.
In addition,
$S \propto (E_g-E_g^*)/T$.
This means that $E_g^*$ defines the boundary 
between electron-like and  hole-like transport processes.
This is in agreement with the fact that $E_g^*$ is in the VFI regime.
Note that the magnitude of $S$ exceeds
$k_{\rm B}/e \simeq 86$ [$\mu$V/K].

%--------------------------------------------
The above results of $G$ and $S$ are explained largely by
$\rho_{\rm d}(\omega)$.
When $U=0$,
\begin{eqnarray} 
\rho_d(\omega)  & = & 
\frac{-1}{\pi} {\rm Im} \frac{1}{\omega - E_g + i \Gamma_0 |\omega/D|}.
\label{eq:rho}
\end{eqnarray}
To adjust the QD level shift due to $U$,
we replace $E_g$ by $E_g-E_g^*$.
When $U=0$, the peak width of $\rho_d(\omega)$, $\Gamma_0 |\omega/D|$, is
smaller than $T$, since $D/\Gamma_0 > 1$ and $T/\Gamma_0 \ll 1$.
This is valid for the present model.
Then, $\rho_d(\omega)$ can be approximated using
a Dirac delta function inside the integral of eq.~(\ref{eq:integral});
$\rho_d(\omega) \sim \delta(\omega-(E_g-E_g^*))$.
This explains the linear dependence of $G$ near the cusp,
and the peak separation of $G$ is proportional to $T$.
Similarly,
$I_1(T)$ in eq.~(\ref{eq:integral}) is proportional to
$|E_g-E_g^*|(E_g-E_g^*)$ near
$E_g=E_g^*$, resulting in $S \propto (E_g-E_g^*)/T$.
The large magnitude of $S$ follows from the fact that
$f'(\omega)$ is finite when $|E_g-E_g^*| \sim T$.

%-------------------------------------------
% Figure: 
%     G vs mu0 and TK vs mu0
%-------------------------------------------
\begin{figure} 
\includegraphics[width=7cm]{%
%./fig3_uni
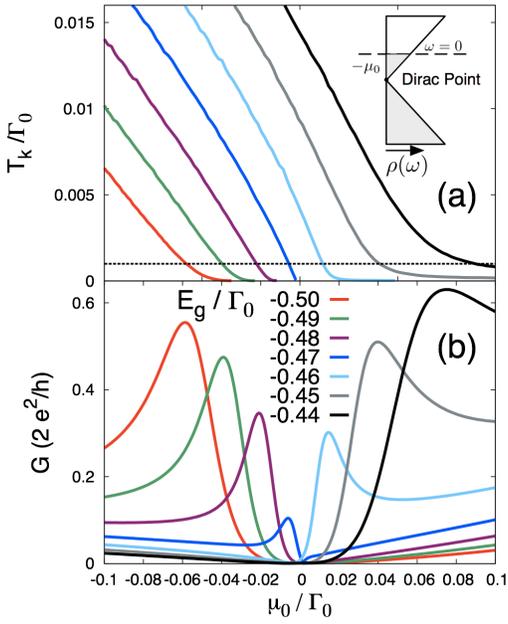}
\caption{%
(Color online)
(a) Kondo temperature $T_K/\Gamma_0$ vs $\mu_0$ for various gate voltages;
$T_K$ is defined by $T_{\rm K}  \chi(T_{\rm K})=0.071$.
The dotted line indicates $T_{\rm K}/\Gamma_0=1.0 \times 10^{-3}$.
Inset: schematic picture of density of states in the leads.
(b) Conductance $G$ vs $\mu_0$ at $T/\Gamma_0 = 1.0 \times 10^{-3}$.
}
\label{fig:mudep}
\end{figure}
%--------------------------------------------
{\it Doped system---}
Next, we consider the case where
the Fermi level ($\omega=0$) is away from the Dirac point,
$\mu_0 \neq 0$,
as depicted in the inset of Fig.~\ref{fig:mudep}(a).
In this case,
a clear sign of the transition between the ASC and LM states appears with the tuning
of the external voltages~\cite{Sengupta:2008gf,Vojta:2010dw}.
We focus on the transition induced by $\mu_0$.
In general, $\mu_0$ determines
the renormalization of the dot level,
and the position of the renormalized level controls
the transition.

In Fig.~\ref{fig:mudep}(a),
the Kondo temperature $T_{\rm K}$ is plotted as a function
of $\mu_0$ for several $E_{\rm g}$ values,
where
$T_{\rm K}$ in this paper is
defined in terms of $\chi$ by
$T_{\rm K} \chi(T_{\rm K}) = 0.0701$~\cite{Krishnamurthy:1980bf,DiasDaSilva:2009gl}.
There is an asymmetry with respect to the sign of $\mu_0$.
In particular,
$T_{\rm K}$ vanishes at $\mu_0=\mu_0^*$ 
when $\mu_0^* < 0$;
$T_{\rm K}=0$ indicates the LM phase.
When $E_{\rm g}$ approaches $E_{\rm g}^*$,
where $E_{\rm g}^*/\Gamma_0\simeq-0.468$ in Fig.~\ref{fig:egdep},
$\mu_0^*$ goes to zero.
This asymmetry has been reported in ref.~\citen{Vojta:2010dw}
for the critical Kondo coupling model, $E_{\rm g}=E_{\rm g}^*$.
A linear dependence of $T_{\rm K}$ is found in
all curves.
It is seen most clearly in the curve of $E_{\rm g}/\Gamma_0=-0.47$ (blue line).
This coincides with the previous results~\cite{Withoff:1990ce,Sengupta:2008gf,Vojta:2010dw}.
The linear coefficient is about 0.368
for $E_{\rm g}=E_{\rm g}^*$~\cite{Withoff:1990ce,Vojta:2010dw},
while here it is about $0.33$ in the vicinity of $E_{\rm g}^*$.
There are deviations from linearity when $T_{\rm K}$ approaches zero
and when $E_g$ is away from $E_g^{*}$. 
We find that the linear coefficient weakly depends on $E_g$.

The behavior of $T_{\rm K}$ correlates with $G$.
Figure~\ref{fig:mudep}(b) shows
$G$ as a function of $\mu_0$ for $T/\Gamma_0=1.0\times 10^{-3}$.
It has a peak structure; at the peak, $T \simeq T_{\rm K}$,
as indicated in (a) by
the intersection of each curve with
the dotted line.
When $T_{\rm K} = 0$, where the system is the LM regime,
$G$ is small.
As $\mu_0$ decreases, $T_{\rm K}$ increases when $\mu_0 < \mu_0^*$.
This results in an increase in $G$ since $T_{\rm K}/T$ increases.
When $\mu_0$ further decreases, $G$ starts to decrease.
This indicates that the system enters the VFI regime
as in a conventional QD since $\mu_0$ controls the effective dot level.
(This point will be further discussed below.)

%-------------------------------------------
% Figure: 
%  G, S, G/G0 vs T, T/TK
%-------------------------------------------
\begin{figure}
\includegraphics[width=9cm]{%
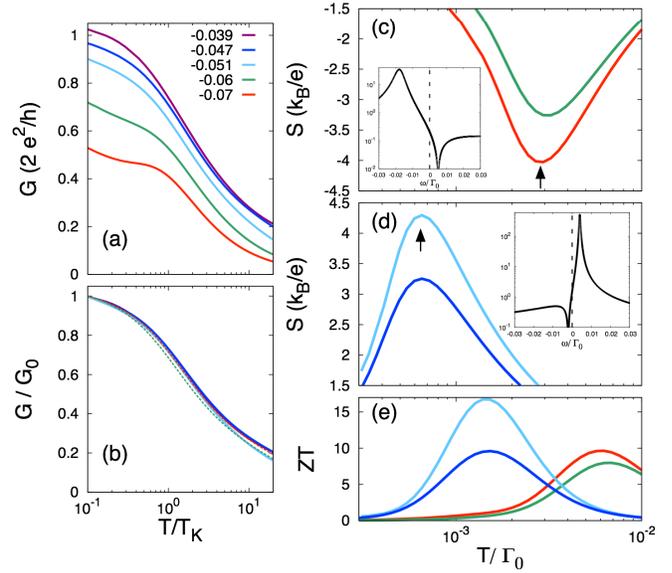}
\caption{%
(Color online)
(a) Conductance $G$ vs $T/T_{\rm K}$ at $E_{\rm g}/\Gamma_0=-0.5$ for various $\mu_0/\Gamma_0$ values.
%as indicated in (a).
(b) Normalized conductance $G/G_0$ vs $T/T_{\rm K}$ with
$G_0=G(0.1T_{\rm k})$
for
three uppermost curves in (a), and
$(E_g/\Gamma_0, \mu_0/\Gamma_0)=$
$(-0.48,-0.015)$ (dotted red line) and
$(-0.47,-1.75\times 10^{-3})$ (dotted green line).
(c,d,e) Temperature dependences of $S$ (c,d) and figure of merit $ZT$ (e)
for
$(E_g/\Gamma_0, \mu_0/\Gamma_0)=$
$(-0.5,-0.005)$(red),
$(-0.5,-0.01)$(green),
$(-0.46,0.002)$(light blue), and
$(-0.46,0.004)$(blue).
Insets: (c,d)
local densities of states [$\rho_{\rm d}(\omega)$] in units of $1/\Gamma_0$
at the points indicated by the arrows.}
\label{fig:temp}
\end{figure}

Now, we discuss the temperature dependence of $G$.
Figure~\ref{fig:temp}(a) shows $G$ as a function of $T/T_K$
for several $\mu_0$ values at $E_g/\Gamma_0 = -0.5$. 
All the curves show that $G$ increases as $T$ decreases.
When $\mu_0$ approaches $\mu_0^*$,
$G$ reaches the unitary limit.
In graphene QD systems,
the Coulomb blockade peak heights are about $0.1 e^2/h$
because of the pseudogap~\cite{%
Stampfer:2008dq,Ponomarenko:2008ej,
2009NanoL...9.2891M,
Guttinger:2010in}.
Thus, this increase in $G$ may provide direct evidence of the Kondo effect
in a graphene QD.
When $\mu_0$ decreases, $G(T)$ shows a hump.
This feature appears in the VFI regime in the conventional QD system
\cite{GoldhaberGordon:1998he,Schoeller:2000dx}.
This means that the crossover from the ASC state to the VFI state appears
when $\mu_0$ changes.
This is consistent with the peak structure of $G$
in Fig.~\ref{fig:mudep}(b).
%The thermopower $S$ changes its sign at $T \simeq T_{\rm K}$,
%which is similar to the conventional Kondo effect
%~\cite{Bickers:1987dc}.
Finally, we comment on the universality of $G$ within the NCA.
In Fig.~\ref{fig:temp}(b), $G/G(T=0.1T_{\rm K})$ is plotted as a function of $T/T_{\rm K}$
in the vicinity of the transition point,
for five different values of $\mu_0$ and $E_g$.
These curves collapse to a universal curve.

Now, we discuss the thermoelectrical properties
in the vicinity of $\mu_0=0$,
where $G \simeq 0$.
Figures~\ref{fig:temp}(c) and~\ref{fig:temp}(d) show the curves of
$S$ plotted as a function of $T$
in the LM (c) and the VFI (d) regimes,
respectively. 
In both cases, $|S| > k_{\rm B}/e$;
$S$ has a negative sign for LM and a positive sign for VFI.
The insets show $\rho_{\rm d}(\omega)$ at the peaks of $S$
as indicated by the arrows.
In those figures,
a sharp peak and a dip appear on opposite sides of
the Fermi level $(\omega=0)$;
the dip originates from the pseudogap and the peak from 
the renormalized localized level.
This asymmetry of $\rho_{\rm d}(\omega)$ leads to the increase in $|S|$ since 
$I_1$ in eq.~(\ref{eq:integral}) is the ``moment'' of
$\rho_{\rm d}(\omega) \Gamma(\omega)$.
We emphasize that the presence of the pseudogap is essential to the increase
since both the sharp peak and the dip originate from the pseudogap.

The increase in $|S|$ indicates that
the QD system
can be a good thermoelectrical device,
which is characterized by
the figure of merit, $ZT=S^2 G T/\kappa$~\cite{Dubi:2011ko},
where $\kappa$ is the thermal conductivity,
given by $\kappa = [I_2(T)-I_1(T)^2/I_0(T)]/T$~\cite{Dong:2002by,2010PhRvB..81x5323L}.
The reason for this is as follows.
Since thermal transport is determined by electron transport,
$\kappa$ is expected to be on the order of $T G$.
[The Wiedemann-Franz law,
$\kappa/(T G) = \pi^2/3 (k_{\rm B}/e)^2$ is not satisfied since the system is not in the Fermi liquid regime.]
This means that $ZT$ can be increased by $S$.
When $ZT > 1$, the system is regarded as a good thermoelectrical material.
Recent studies
\cite{Xia:2009fn,Hsieh:2009dp,Hsieh:2009ig,Chen:2009do,Zhang:2009ks}
showed that 
a single Dirac cone appears on the surfaces of
Bi$_2$Te$_3$, Bi$_2$Se$_3$, and Sb$_2$Te$_3$.
Those materials are also known as good thermoelectrical materials.
Figure~\ref{fig:temp}(e) shows $ZT$ as a function of $T$ in the LM and VFI regimes.
This figure also shows  that $ZT > 1$ is achieved when $|S| > k_{\rm B}/e$;
moreover, $ZT \gtrsim 10$ in a certain range.
The results show that the increase in $ZT$ correlates with the increase in $|S|$
as explained above.
Our results are also consistent with
the theory on the optimization of $ZT$
by Mahan and Sofo~\cite{Mahan23071996},
where 
a narrow peak distribution of conduction electrons increases $ZT$.
In the present case, the pseudogap induces the narrow peak as shown in
the insets of
Figs.~\ref{fig:temp}(c) and (d).

%%%%%%%%%%%%%%%%%%%%%%%%%%%%%%%%%%%%
% CONCLUSIONS
%%%%%%%%%%%%%%%%%%%%%%%%%%%%%%%%%%%%

In summary, we have investigated 
electrical and thermoelectrical properties of 
the pseudogap Anderson model.
In the undoped case,
the conductance is much less than $2e^2/h$,
whereas the magnitude of the thermopower exceeds $k_{\rm B}/e$.
In the doped case, the conductance shows a peak structure,
which indicates the transition between the ASC and LM phases,
characterized by the Kondo temperature.
The unitary limit of the conductance appears at low temperatures.
When the magnitude of the thermopower exceeds $k_{\rm B} /e$,
the figure of merit is greater than unity.
The QD system in Dirac fermions therefore
displays two distinctive features induced by the pseudogap,
an impurity quantum phase transition,
and excellent thermoelectrical properties.

\begin{acknowledgments}
The author thanks 
Y.~Avishai, A.~Golub, T.~Nakanishi, and Y.~Takane 
for fruitful discussion and comments
on various aspects.
The author also acknowledges the support from JSPS and JST.

\end{acknowledgments}

%\bibliographystyle{jpsj} 
%\bibliography{Outline222}

\begin{thebibliography}{10}

\bibitem{Neto:2009cl}
A.~H. Castro~Neto, F.~Guinea, N.~M.~R. Peres, K.~S. Novoselov, and A.~K. Geim:
  Rev. Mod. Phys. {\bfseries 81} (2009) 109.

\bibitem{Stampfer:2008dq}
C.~Stampfer, J.~G{\"u}ttinger, F.~Molitor, D.~Graf, T.~Ihn, and K.~Ensslin:
  App. Phys. Lett. {\bfseries 92} (2008) 012102.

\bibitem{Ponomarenko:2008ej}
L.~A. Ponomarenko, F.~Schedin, M.~I. Katsnelson, R.~Yang, E.~W. Hill, K.~S.
  Novoselov, and A.~K. Geim: Science {\bfseries 320} (2008) 356.

\bibitem{2009NanoL...9.2891M}
S.~Moriyama, D.~Tsuya, E.~Watanabe, S.~Uji, M.~Shimizu, T.~Mori, T.~Yamaguchi,
  and K.~Ishibashi: Nano Lett. {\bfseries 9} (2009) 2891.

\bibitem{Guttinger:2010in}
J.~G{\"u}ttinger, T.~Frey, C.~Stampfer, T.~Ihn, and K.~Ensslin: Phys. Rev.
  Lett. {\bfseries 105} (2010) 116801.

\bibitem{1998Natur.391..156G}
D.~Goldhaber-Gordon, H.~Shtrikman, D.~Mahalu, D.~Abusch-Magder, U.~Meirav, and
  M.~A. Kastner: Nature {\bfseries 391} (1998) 156.

\bibitem{1998Sci...281..540C}
S.~M. Cronenwett, T.~H. Oosterkamp, and L.~P. Kouwenhoven: Science {\bfseries
  281} (1998) 540.

\bibitem{Sengupta:2008gf}
K.~Sengupta and G.~Baskaran: Phys. Rev. B {\bfseries 77} (2008) 045417.

\bibitem{Uchoa:2009hz}
B.~Uchoa, L.~Yang, S.~W. Tsai, N.~M.~R. Peres, and A.~H. Castro~Neto: Phys.
  Rev. Lett. {\bfseries 103} (2009) 206804.

\bibitem{Cornaglia:2009ir}
P.~S. Cornaglia, G.~Usaj, and C.~A. Balseiro: Phys. Rev. Lett. {\bfseries 102}
  (2009) 046801.

\bibitem{Zhuang:2009ji}
H.-B. Zhuang, Q.-f. Sun, and X.~C. Xie: Europhys. Lett. {\bfseries 86} (2009)
  58004.

\bibitem{Wehling:2010in}
T.~O. Wehling, A.~V. Balatsky, M.~I. Katsnelson, A.~I. Lichtenstein, and
  A.~Rosch: Phys. Rev. B {\bfseries 81} (2010) 115427.

\bibitem{Vojta:2010dw}
M.~Vojta, L.~Fritz, and R.~Bulla: Europhys. Lett. {\bfseries 90} (2010) 27006.

\bibitem{Uchoa:2011gt}
B.~Uchoa, T.~G. Rappoport, and A.~H. Castro~Neto: Phys. Rev. Lett. {\bfseries
  106} (2011) 016801; {\bfseries
  106} (2011) 159901(E).

%\bibitem{Uchoa:2011fr}
%B.~Uchoa, T.~G. Rappoport, and A.~H. Castro~Neto: Phys. Rev. Lett. {\bfseries
%  106} (2011) 159901(E).

\bibitem{2012arXiv1208.3113F}
L.~{Fritz} and M.~{Vojta}:
%arXiv:1208.3113  (2012).
Rep.~Prog.~Phys. {\bfseries 76} (2013) 032501.


\bibitem{Chen:2011jm}
J.-H. Chen, L.~Li, W.~G. Cullen, E.~D. Williams, and M.~S. Fuhrer: Nat. Phys.
  {\bfseries 7} (2011) 535.

\bibitem{Xia:2009fn}
Y.~Xia, D.~Qian, D.~Hsieh, L.~Wray, A.~Pal, H.~Lin, A.~Bansil, D.~Grauer, Y.~S.
  Hor, R.~J. Cava, and M.~Z. Hasan: Nat. Phys. {\bfseries 5} (2009) 398.

\bibitem{Hsieh:2009dp}
D.~Hsieh, Y.~Xia, D.~Qian, L.~Wray, J.~H. Dil, F.~Meier, J.~Osterwalder,
  L.~Patthey, J.~G. Checkelsky, N.~P. Ong, A.~V. Fedorov, H.~Lin, A.~Bansil,
  D.~Grauer, Y.~S. Hor, R.~J. Cava, and M.~Z. Hasan: Nature {\bfseries 460}
  (2009) 1101.

\bibitem{Hsieh:2009ig}
D.~Hsieh, Y.~Xia, D.~Qian, L.~Wray, F.~Meier, J.~H. Dil, J.~Osterwalder,
  L.~Patthey, A.~V. Fedorov, H.~Lin, A.~Bansil, D.~Grauer, Y.~S. Hor, R.~J.
  Cava, and M.~Z. Hasan: Phys. Rev. Lett. {\bfseries 103} (2009) 146401.

\bibitem{Chen:2009do}
Y.~L. Chen, J.~G. Analytis, J.-H. Chu, Z.~K. Liu, S.-K. Mo, X.~L. Qi, H.~J.
  Zhang, D.~H. Lu, X.~Dai, Z.~Fang, S.~C. Zhang, I.~R. Fisher, Z.~Hussain, and
  Z.-X. Shen: Science {\bfseries 325} (2009) 178.

\bibitem{Zhang:2009ks}
H.~Zhang, C.-X. Liu, X.-L. Qi, X.~Dai, Z.~Fang, and S.-C. Zhang: Nat. Phys.
  {\bfseries 5} (2009) 438.

\bibitem{Liu:2009ig}
Q.~Liu, C.-X. Liu, C.~Xu, X.-L. Qi, and S.-C. Zhang: Phys. Rev. Lett.
  {\bfseries 102} (2009) 156603.

\bibitem{Tran:2010dw}
M.-T. Tran and K.-S. Kim: Phys. Rev. B {\bfseries 82} (2010) 155142.

\bibitem{Zitko:2010bp}
R.~{\v Z}itko: Phys. Rev. B {\bfseries 81} (2010) 241414.

\bibitem{Feng:2010jx}
X.-Y. Feng, W.-Q. Chen, J.-H. Gao, Q.-H. Wang, and F.-C. Zhang: Phys. Rev. B
  {\bfseries 81} (2010) 235411.

\bibitem{2012arXiv1211.0034M}
A.~K. {Mitchell}, D.~{Schuricht}, M.~{Vojta}, and L.~{Fritz}: 
%arXiv:1211.0034 (2012).
Phys. Rev. B {\bfseries 87} (2013) 075430.

\bibitem{Withoff:1990ce}
D.~Withoff and E.~Fradkin: Phys. Rev. Lett. {\bfseries 64} (1990) 1835.

\bibitem{Cassanello:1996jg}
C.~R. Cassanello and E.~Fradkin: Phys. Rev. B {\bfseries 53} (1996) 15079;
 {\bfseries 56} (1997) 11246.

%\bibitem{Cassanello:1997ku}
%C.~R. Cassanello and E.~Fradkin: Phys. Rev. B {\bfseries 56} (1997) 11246.

\bibitem{GonzalezBuxton:1996cz}
C.~Gonzalez-Buxton and K.~Ingersent: Phys. Rev. B {\bfseries 54} (1996) R15614;
{\bfseries 57} (1998) 14254.

%\bibitem{GonzalezBuxton:1998hb}
%C.~Gonzalez-Buxton and K.~Ingersent: Phys. Rev. B {\bfseries 57} (1998) 14254.

\bibitem{Bulla:1997fp}
R.~Bulla, {\relax Th}.~Pruschke, and A.~C. Hewson: J. Phys. Condens. Matter
  {\bfseries 9} (1997) 10463.

\bibitem{Fritz:2004cn}
L.~Fritz and M.~Vojta: Phys. Rev. B {\bfseries 70} (2004) 214427.

\bibitem{Vojta:2004gn}
M.~Vojta and L.~Fritz: Phys. Rev. B {\bfseries 70} (2004) 094502.

\bibitem{Kanao:2012ho}
T.~Kanao, H.~Matsuura, and M.~Ogata: J. Phys. Soc. Jpn. {\bfseries 81} (2012)
  063709.

\bibitem{Dubi:2011ko}
Y.~Dubi and M.~Di~Ventra: Rev. Mod. Phys. {\bfseries 83} (2011) 131.

\bibitem{Kondo:1965gu}
J.~Kondo: Prog. Theor. Phys. {\bfseries 34} (1965) 372.

\bibitem{2001EL.....56..576B}
D.~Boese and R.~Fazio: Europhys. Lett. {\bfseries 56} (2001) 576.

\bibitem{Dong:2002by}
B.~Dong and X.~L. Lei: J. Phys. Condens. Matter {\bfseries 14} (2002) 11747.

\bibitem{Scheibner:2005kh}
R.~Scheibner, H.~Buhmann, D.~Reuter, M.~N. Kiselev, and L.~W. Molenkamp: Phys.
  Rev. Lett. {\bfseries 95} (2005) 176602.

\bibitem{Coleman:1984fz}
P.~Coleman: Phys. Rev. B {\bfseries 29} (1984) 3035.

\bibitem{Bickers:1987dc}
N.~E. Bickers: Rev. Mod. Phys. {\bfseries 59} (1987) 845.

\bibitem{Wingreen:1994hw}
N.~S. Wingreen and Y.~Meir: Phys. Rev. B {\bfseries 49} (1994) 11040.

\bibitem{Vojta:2001fo}
M.~Vojta: Phys. Rev. Lett. {\bfseries 87} (2001) 097202.

\bibitem{note1}
For the $K'$ point, the off-diagonal elements of $M$ are
replaced by their complex conjugates.

\bibitem{Barnes:1976tq}
S.~E. Barnes: J. Phys. F {\bfseries 6} (1976) 1375; {\bfseries 7} (1977) 2637.

%\bibitem{Barnes:1977um}
%S.~E. Barnes: J. Phys. F {\bfseries 7} (1977) 2637.

\bibitem{2010PhRvB..81x5323L}
J.~Liu, Q.-f. Sun, and X.~C. Xie: Phys. Rev. B {\bfseries 81} (2010) 245323.

\bibitem{Krishnamurthy:1980bf}
H.~R. Krishna-murthy, J.~W. Wilkins, and K.~G. Wilson: Phys. Rev. B {\bfseries
  21} (1980) 1003.

\bibitem{DiasDaSilva:2009gl}
L.~G. G.~V. Dias~da Silva, N.~Sandler, P.~Simon, K.~Ingersent, and S.~E. Ulloa:
  Phys. Rev. Lett. {\bfseries 102} (2009) 166806.

\bibitem{GoldhaberGordon:1998he}
D.~Goldhaber-Gordon, J.~G{\"o}res, M.~A. Kastner, H.~Shtrikman, D.~Mahalu, and
  U.~Meirav: Phys. Rev. Lett. {\bfseries 81} (1998) 5225.

\bibitem{Schoeller:2000dx}
H.~Schoeller and J.~K{\"o}nig: Phys. Rev. Lett. {\bfseries 84} (2000) 3686.


\bibitem{Mahan23071996}
G.~D.~Mahan and J.~O.~Sofo:
Proc.~Natl.~Acad.~Sci.~U.~S.~A.
{\bfseries 93} (1996) 7436.

\end{thebibliography}

\providecommand{\noopsort}[1]{}\providecommand{\singleletter}[1]{#1}%

%\begin{thebibliography}{9}
%\bibitem{jpsj} The abbreviation for JPSJ must be ``J. Phys. Soc. Jpn." \note{in the %reference list}.
%\bibitem{instructions} More abbreviations of journal titles are listed in ``Instructions for Preparation of Manuscript".
%\end{thebibliography}

\end{document}